\documentclass[useAMS]{mn2e}
\usepackage{graphicx}
\usepackage{multirow}
\usepackage{txfonts}
\usepackage{color}
\usepackage{amssymb}
\usepackage[T1]{fontenc}
\usepackage{aecompl}

\title[Interpretation of the $\nu$ Eri oscillations: a call for modification of stellar opacities]
{Interpretation of the BRITE oscillation data
 of the hybrid pulsator $\nu$\,Eridani: a call for the modification of stellar opacities}
\author[Daszy\'nska-Daszkiewicz et al.]{ J. Daszy\'nska-Daszkiewicz$^{1}$\thanks{E-mail:daszynska@astro.uni.wroc.pl},
A. A. Pamyatnykh$^{2}$, P. Walczak$^{1}$, J. Colgan$^{3}$,
\newauthor  C. J. Fontes$^{3}$, D. P. Kilcrease$^{3}$ \\
$^{1}$Instytut Astronomiczny, Uniwersytet Wroc{\l}awski, Kopernika 11, 51-622 Wroc{\l}aw, Poland\\
$^{2}$Nicolaus Copernicus Astronomical Center, Bartycka 18, 00-716, Warsaw, Poland\\
$^{3}$Los Alamos National Laboratory, Los Alamos, NM 87545, USA\\
}

\begin{document}

\date{Accepted 1988 December 15. Received 1988 December 14; in original form 1988 October 11}

\pagerange{\pageref{firstpage}--\pageref{lastpage}} \pubyear{2002}

\maketitle

\label{firstpage}

\begin{abstract}
The analysis of the BRITE oscillation spectrum of the main sequence early B-type star $\nu$ Eridani is presented.
Only models with the modified mean opacity profile can account for the observed frequency ranges as well as for the values
of some individual frequencies.
The number of the $\kappa$ modified seismic models is constrained by the nonadiabatic parameter $f$,
which is very sensitive to the opacity changes in the subphotospheric layers where the pulsations are driven.
We present an example of the model that satisfies all the above conditions.
It seems that the OPLIB opacities are preferred over those from the OPAL and OP projects.
Moreover, we discuss additional consequences of the opacity modification, namely,
an enhancement of the efficiency of convection in the Z-bump as well as an occurrence of close radial modes
which is a kind of avoided-crossing phenomenon common for nonradial modes in standard main sequence models.
\end{abstract}

\begin{keywords}
stars: early-type -- stars: oscillations -- stars: individual: $\nu$ Eri -- atomic data: opacities
\end{keywords}

\section{Introduction} \label{introduction}

One of the main ingredients in stellar modelling are opacity data which affect energy transport and the structure of a star.
In addition, the values of opacities determine conditions for excitation of heat-driven pulsations as observed, e.g., in main-sequence
or classical pulsators.

The profound consequences that can result from the use of incorrect opacities was demonstrated some quarter of a century ago when the new data
were computed
by the two independent teams: OPAL (Iglesias, Rogers \& Wilson 1992, Rogers \& Iglesias 1992) and OP (Seaton 1993, Seaton et al. 1994).
Since then, these opacity data were updated several times (Iglesias \& Rogers 1996, Seaton 2005), but the main feature, i.e.,
the new local maximum of the Rosseland mean opacity, $\kappa$,
at the temperature $T\approx 200~000$ K, was identified in their first release.
This new opacity bump, called the metal or Z-bump, made possible, e.g., to explain pulsations of B-type main sequence stars,
decrease the mass discrepancy in classical Cepheids and to improve the standard solar model (Rogers \& Iglesias 1994).
Recently, the Los Alamos opacity database (OPLIB) was also updated (Colgan et al. 2015, 2016) and made publicly available
(http://aphysics2.lanl.gov/cgi-bin/opacrun/astro.pl).
Calculations with the OPLIB data yielded wider instability strips for pulsations of $\beta$ Cep and SPB type (Walczak et al. 2015).

The most recent laboratory measurements at solar interior temperatures (Bailey et al. 2015) indicate that the predicted
Rosseland mean opacities for iron are underestimated by about 75\%.
Despite various improvements in the calculation of stellar opacities, there are many indications that something is still missing
in these data and/or has not been correctly included. One example is the B-type main sequence pulsators, which exhibit both pressure
and high-order gravity modes, e.g., $\nu$ Eridani (Handler et al. 2004, Aerts et al. 2004), 12 Lacertae (Handler et al. 2006),
$\gamma$ Pegasi (Handler et al. 2009).
There are also bags of such pulsators identified
from the space data from \emph{Kepler} and \emph{CoRoT} missions (e.g., Degroote et al. 2009, Balona et al. 2011, 2015a).
However, so far none of the pulsational models can account for their oscillation spectra.

The higher iron opacity has been investigated very recently by theoretical computations of Nahar \& Pradhan (2016),
but their results regarding a potential increase in total opacity have been criticized (Blancard et al. 2016).
Larger opacities could be a breakthrough in solar modelling and push in the right direction the seismic studies of massive stars.
The increase of the Z-bump opacity, mostly dominated by iron, was a first guess to explain the low frequencies detected
in $\nu$ Eri (Pamyatnykh, Handler \& Dziembowski 2004). Recently, the instability strips for B-type main sequence models
with enhancement by 75 \% iron and nickel opacity were published by Moravveji (2016).

In this paper, we make an attempt to analyse once again this well-known hybrid pulsator, $\nu$ Eri,
which was observed by the BRITE-Constellation (Weiss, Ruci\'nski, Moffat et al. 2014).
Here, we base our studies on the pulsational frequencies extracted by Handler et al. (2016) from the BRITE light curves.
Our goal is to modify the profile of the mean opacity to reproduce both the observed range of frequencies and
the values of the frequencies themselves. Moreover, in order to control the opacity changes and make our results more plausible,
we have imposed a requirement for the relative amplitude of the bolometric flux variations, the so-called parameter $f$
(Cugier, Dziembowski, Pamyatnykh 1994, Daszy\'nska-Daszkiewicz et al. 2002).
Namely, we demand, in addition, that the empirical value of $f$ for the dominant radial mode is reproduced by a model
satisfying the above conditions. The use of the parameter $f$ significantly limits the modification of opacities
because this parameter is very sensitive to the structure of subphotospheric layers where the pulsation driving occurs.

In Sect.\,2, we recall the observed properties of pulsations of $\nu$ Eri and compare them with the standard pulsational models.
By the standard models we mean the models computed with the available opacity data: OPLIB, OPAL and OP.
Sect.\,3 contains results of our modelling extended by the modification of the mean opacity profile and in Sect.\,4
we reduce the number of such seismic models by taking into account the parameter $f$.
The effect of convection in the vicinity of the metal (or $Z-$)bump on pulsational properties of models with masses typical for $\nu$ Eri
is discussed in Sect.\,5. A kind of avoided-crossing phenomenon for radial modes identified in models with
the modified $\kappa$ profile
is presented in Sect.\,6. We end with Conclusions and perspectives for future works.

\section{Oscillations of $\nu$ Eri and the standard pulsational models} \label{hd163899}

The early B-type star $\nu$ Eri (B2III, HD 29248) is one of the most studied hybrid pulsators in the last fifteen years.
These intensive studies began with dedicated photometric and spectroscopic multisite campaigns carried out in 2003-2005
(Handler et al. 2004, Jerzykiewicz et al. 2005, Aerts et al. 2004, de Ridder et al. 2004). The analysis of these data enriched
the known oscillation spectrum by new frequency peaks, that complemented the two known $\ell=1$ triplets, as well as revealed
the existence of high order g modes associated with low frequencies.
Subsequently, many theoretical studies have been devoted to finding the best seismic models that yield constraints
on overshooting from the convective core, internal rotation and opacities (Pamyatnykh, Handler \& Dziembowski 2004,
Ausseloos et al. 2004, Daszy\'nska-Daszkiewicz, Dziembowski, Pamyatnykh 2005, Dziembowski \& Pamyatnykh 2008, Suarez et al. 2009,
Daszy\'nska-Daszkiewicz \& Walczak 2010).
This star is a slow rotator with $V_{\rm rot}=6$ km/s (Pamyatnykh, Handler \& Dziembowski 2004)
but some suggestions about non-rigid rotation have been made (Dziembowski \& Pamyatnykh 2008, Suarez et al. 2009).

The BRITE project provided an opportunity, for the first time, to register the light variations of this bright star ($V=3.92$ mag)
from space. Handler et al. (2016) extracted from these data 17 frequencies: 10 in the high frequency range (p modes) and 7 in the low frequency
range (g modes).
Most of these peaks agree with the values derived from the previous photometric and spectroscopic observations. However, there are
a few differences.
Firstly, five more g-mode frequencies have been detected. Interestingly, the "old" frequency $\nu=0.61440$ d$^{-1}$
is missing in this low frequency "forest". Moreover, two frequencies, one component of the $\ell=1,~p_1$ triplet around 6.22 d$^{-1}$
and the peak around $\nu=6.73$ d$^{-1}$, have not been found. These two frequencies had the lowest photometric amplitudes
in the last photometric campaigns (cf. Table 6 in  Jerzykiewicz et al. 2005).
From a theoretical point of view, such a situation is quite plausible because the e-folding time for the amplitude growth of some
pulsational modes in main sequence models may be quite short.
For the quadrupole g mode with $\nu\approx0.61$ d$^{-1}$, we obtained an e-folding time of about 15 years and
for the dipole and quadrupole p modes with $\nu\approx6.22$ and $6.73$ d$^{-1}$, this time is of the order of several decades. These estimates, besides
the detection threshold, can explain the appearance and disappearance of some frequency peaks. It is worth mentioning that
the Str\"omgren $u$ amplitude of the high order g mode $\nu\approx 0.433$ d$^{-1}$ has dropped from about 5.5 mmag
(Handler et al. 2004) to 3.0 mmag (Handler et al. 2016), ie., by a factor of almost two within 12 years.

In Fig.\,1, we show the oscillation spectrum of $\nu$ Eri as obtained by Handler et al. (2016) from the BRITE light curves
(the bottom panel) as well as, for comparison, the oscillation spectrum from the 2002-2004 photometric campaigns (the top panel)
derived by Jerzykiewicz et al. (2005).

\begin{figure}
\includegraphics[width=\columnwidth,clip]{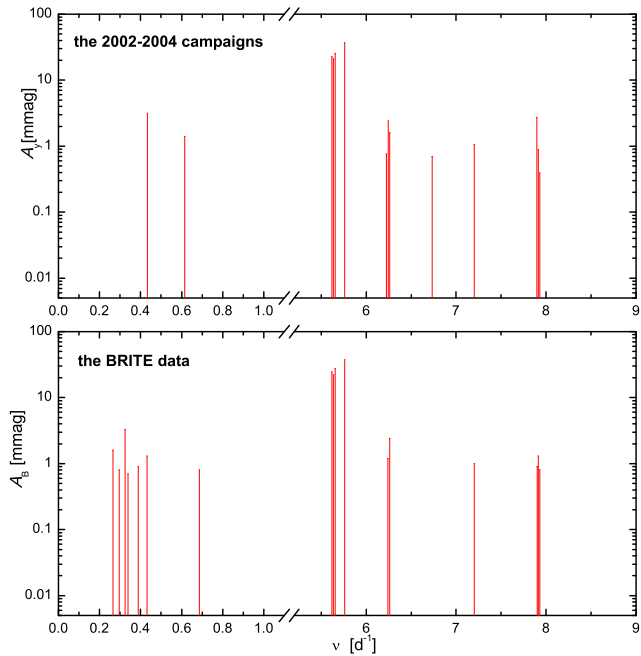}
\caption{The oscillation spectra of $\nu$ Eri obtained from the 2002-2004 photometric campaigns by Jerzykiewicz et al. (2005) (the top panel)
and from the BRITE space photometry by Handler et al. (2016) (the bottom panel). To highlight the peaks,
different scales are used for the low- and high-frequency peaks, and the X-axis is broken in the range
of 1.2-5.2 [d$^{-1}$].}
\label{fig1}
\end{figure}

As a first step, we confront the observed frequency range with the theoretical results obtained from models computed with the standard
opacity data commonly used in pulsational computations: OPAL (Iglesias \& Rogers 1996), OP (Seaton 2005) and the recently released OPLIB library
(Colgan et al. 2015, 2016). In all cases, the solar chemical mixture was adopted (Asplund et al. 2009).
There are some differences in the run of the mean opacity profiles, $\kappa(T)$, and its temperature derivative, $\kappa_T(T)$,
between these three sources of the opacity data. A comparison of them for a typical $\beta$ Cep model was recently presented
by Walczak et al. (2015) (see their Fig.\,1 and 2).

In Fig.\,2, we plot the normalized instability parameter, $\eta$, as a function of the mode frequency for the three models
suitable for $\nu$ Eri computed with the three opacity data sets: OPLIB, OPAL, OP. The unstable modes have positive values of $\eta$.
There are shown modes with the spherical harmonic degree up to $\ell=2$. The effective temperature and luminosity
($\log T_{\rm eff},~\log L/L_\odot$) of these models are consistent with the observed error box determined with
the new parallax $\pi=4.83\pm0.19$ mas (van Leeuwen 2007): $\log T_{\rm eff}=4.346\pm0.014$, $\log L/L_\odot=3.886\pm0.044$.
Here, we adopted a mass $M=9.5M_{\odot}$ and metallicity $Z=0.015$. The values of ($\log T_{\rm eff},~\log L/L_\odot$) differ
only slightly between the models shown in Fig.\,2 and their approximate values are: $\log T_{\rm eff}\approx 4.343$, $\log L/L_\odot\approx 3.92$.
All models reproduce
the radial fundamental mode $\ell=0,~p_1$ and the centroid of the dipole mode $\ell=1,~g_1$ corresponding to the frequencies
$\nu=5.76326$ d$^{-1}$ and $\nu=5.63725$ d$^{-1}$, respectively. A small amount of core overshooting was needed to adjust the dipole mode
frequency, which was chosen from the range $\alpha_{\rm ov}\in (0.07-0.09)$. Fig.\,2 recalls the old problem of mode excitation
in the low frequency range, i.e., high order g modes, as well as in the highest frequency range for $\nu\gtrsim 7.5$ d$^{-1}$
(e.g. Pamyatnykh, Handler \& Dziembowski, 2004).
The widest instability in the p mode range is in the OPLIB models, reaching $\nu\approx 8$ d$^{-1}$,
whereas for the OP models we get the highest values of $\eta$ in the range of high order g modes.
As we have checked, the changes of various parameters (e.g., $M,~Z,~X$) within a range allowed by the observational error box do not eliminate these
shortcomings. Increasing the total metallicity, $Z$, enhanced an overall instability in both local maxima of $\eta$.
The effect of changing the mass is as follows: for higher masses we get the higher instability for the highest frequency p modes,
e.g., changing the mass from 9.0 to 10.0 $M_\odot$ extends the instability by about 1 d$^{-1}$ towards higher frequencies. On the other hand,
for the higher mass models we get lower instability for low frequency g modes.
Thus, we can say that in the allowed range of parameters there are no models that can account for the whole observed range of frequencies
detected in the $\nu$ Eri light variations.
In order to overcome this problem in the next section we try to modify the opacity profile, $\kappa(T)$,
and to find models that reproduce both the observed range of frequencies and the values of some individual frequencies.

\begin{figure}
 \includegraphics[clip,width=\columnwidth]{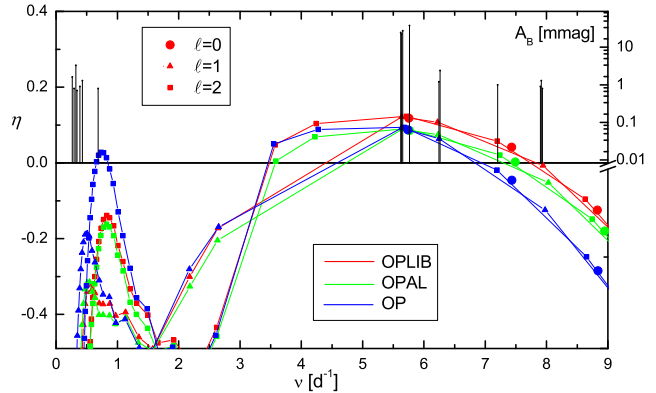}
   \caption{The normalized instability parameter, $\eta$, for representative models of $\nu$ Eri, computed with
the three sources of opacity data: OPLIB, OPAL and OP. All models have $Z=0.015$, $M=9.5~M_\odot$,
$\log T_{\rm eff}\approx 4.343$, $\log L/L_\odot\approx 3.92$ and the overshooting parameter from the range
$\alpha_{\rm ov}\in (0.07-0.09)$. }
\label{fig2}
\end{figure}

\section{Models with modified opacities} \label{models}

In this section we change some amount of opacity at the depths (expressed in $\log T$) that can affect the driving of pulsations
in the stellar models of $\nu$ Eri. Of course, these $\kappa$ modifications will affect also the values of eigenfrequencies.
The opacity profile is modified according to the formula
$$\kappa (T)=\kappa_0(T) \left[1+\sum_{i=1}^N b_i \cdot\exp\left( -\frac{(\log T-\log T_{0,i})^2}{a_i^2}\right) \right],$$
where $\kappa_0(T)$ is the unchanged opacity profile and $(a,~b,~T_0)$ are parameters of a Gaussian describing
the width, height and position of the maximum, respectively. Thus, for a fixed depth, $T_0$, we enhance or reduce
the opacity by changing $(a,~b)$. We adopted the Gaussian because the known local maxima of the mean opacity profile
are well represented by this function.
Examples of modifications of the OPLIB values of $\kappa(\log T)$ are depicted in the top panels of Figs.\,3 and 4
for a model with the parameters $M = 9.5~M_{\sun}$, $\log T_{\rm eff} = 4.3399$, $\log L/L_{\sun} = 3.917$,
metallicity $Z = 0.015$ and core overshooting $\alpha_{\rm ov} = 0.07$.
The $\kappa$ derivative with respect to temperature, $\kappa_T=\partial\log\kappa/\partial\log T$, is also shown,
with its values given on the right-hand Y-axis.
In the first case, we increased opacities around $\log T_0=5.3$ and 5.46 by 100\% (i.e. b=1),
whereas in the second case (the top panel of Fig.\,4), we added also 100\%  opacity around $\log T_0=5.06$.
Within the Z-bump, we have at $\log T_0=5.3$ the maximum contribution of iron to the opacity whereas
at $\log T_0=5.46$ the maximum contribution of nickel occurs (e.g., Salmon et al. 2012).
Adding 100\% of the opacity only at $\log T_0=5.3$ increases the instability of both p- and high order g-mode frequencies
whereas an additional increase of $\kappa$ at the deeper temperature $\log T_0=5.46$ raises the parameter $\eta$ mostly
for the low frequency modes.
The opacity changes at $\log T_0=5.06$ are motivated by works of Cugier (2012, 2014), who identified the new opacity bump
around this temperature in the Kurucz atmosphere models. This bump was suggested, for example, as a possible cause of excitation
of low frequency g modes in $\delta$ Scuti stars as detected in the Kepler data (Balona, Daszy\'nska-Daszkiewicz \& Pamyatnykh 2015).

The consequences of such $\kappa$ modifications on pulsational properties are demonstrated in the middle panels of Figs.\,3 and 4.
As an example we show the normalized differential work integral for the radial second overtone mode ($\ell=0$, p$_3$).
In the model computed with the standard OPLIB data this mode is stable. Increasing the opacity at $\log T=5.3$ and 5.46 makes
this radial mode unstable
whereas adding the third bump at $\log T=5.06$ stabilizes it again. In Fig.\,3, we compare also the work integral for the high order
quadrupole mode $\ell=2$, g$_{16}$ for which
the parameter $\eta$ reaches the maximum.  In the model computed with the standard $\kappa$ profile this mode is stable whereas
in the model computed with the modified $\kappa$ profile
at $\log T=5.3$ and 5.46 it becomes unstable.

The bottom panels of Figs.\,3 and 4 show the run of the normalized instability parameter, $\eta$, for the two considered cases.
As one can see, an enhancement of the opacity at the depths $\log T_0=5.3$ and 5.46 causes an increase of the pulsational
instability both for low and high frequency modes. In the low frequency range, the quadrupole modes become unstable and dipole
modes are not far from the instability. For the high frequencies, the instability range becomes much wider than the observed
frequency range.
If we add the third bump at the depth $\log T_0=5.06$ (Fig.\,4), the increase of $\eta$ for the highest frequency modes is
smaller whereas for low frequency modes the parameter $\eta$ is almost unchanged.

\begin{figure}
 \includegraphics[clip,width=\columnwidth]{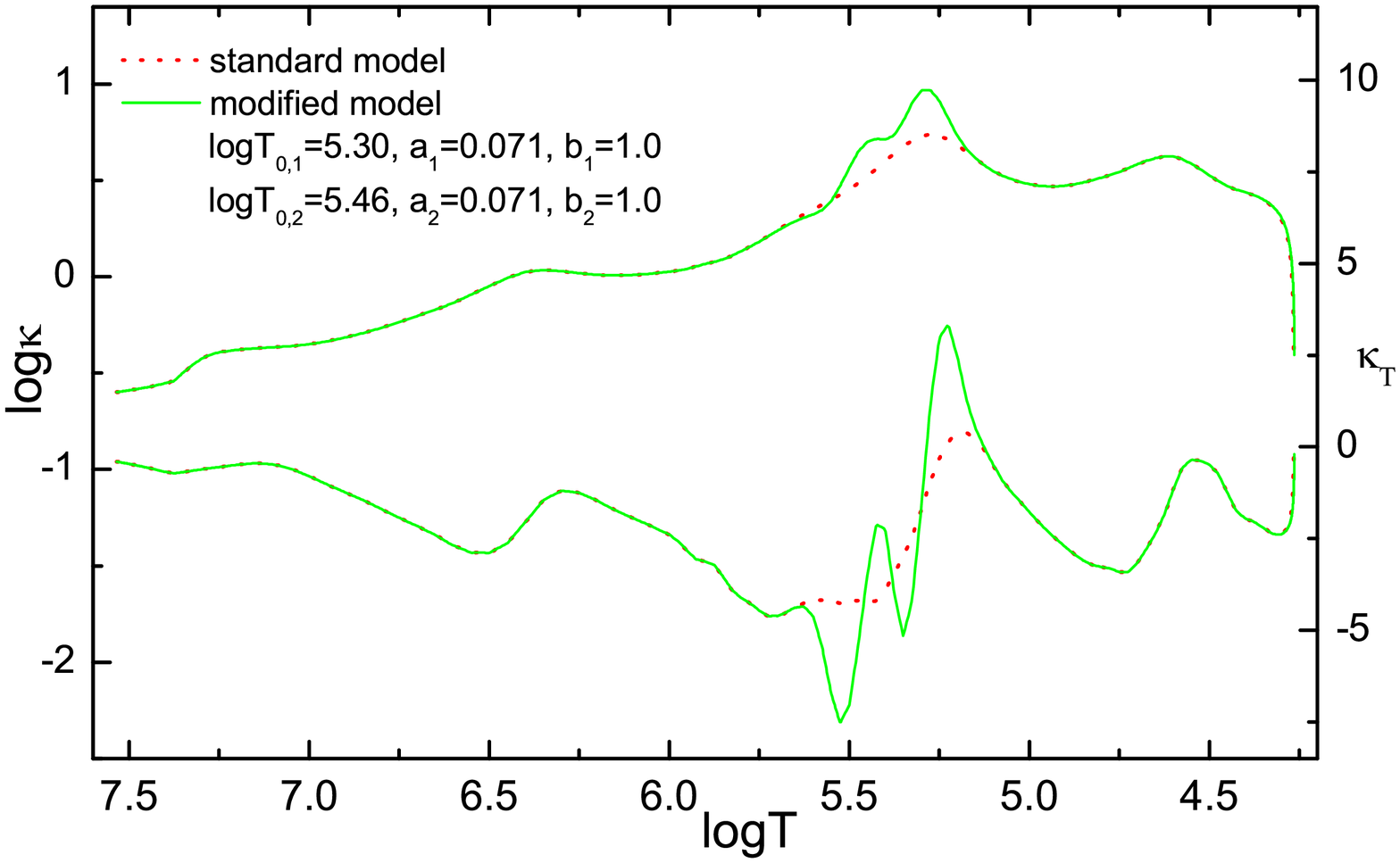}
 \includegraphics[clip,width=\columnwidth,height=4.8cm]{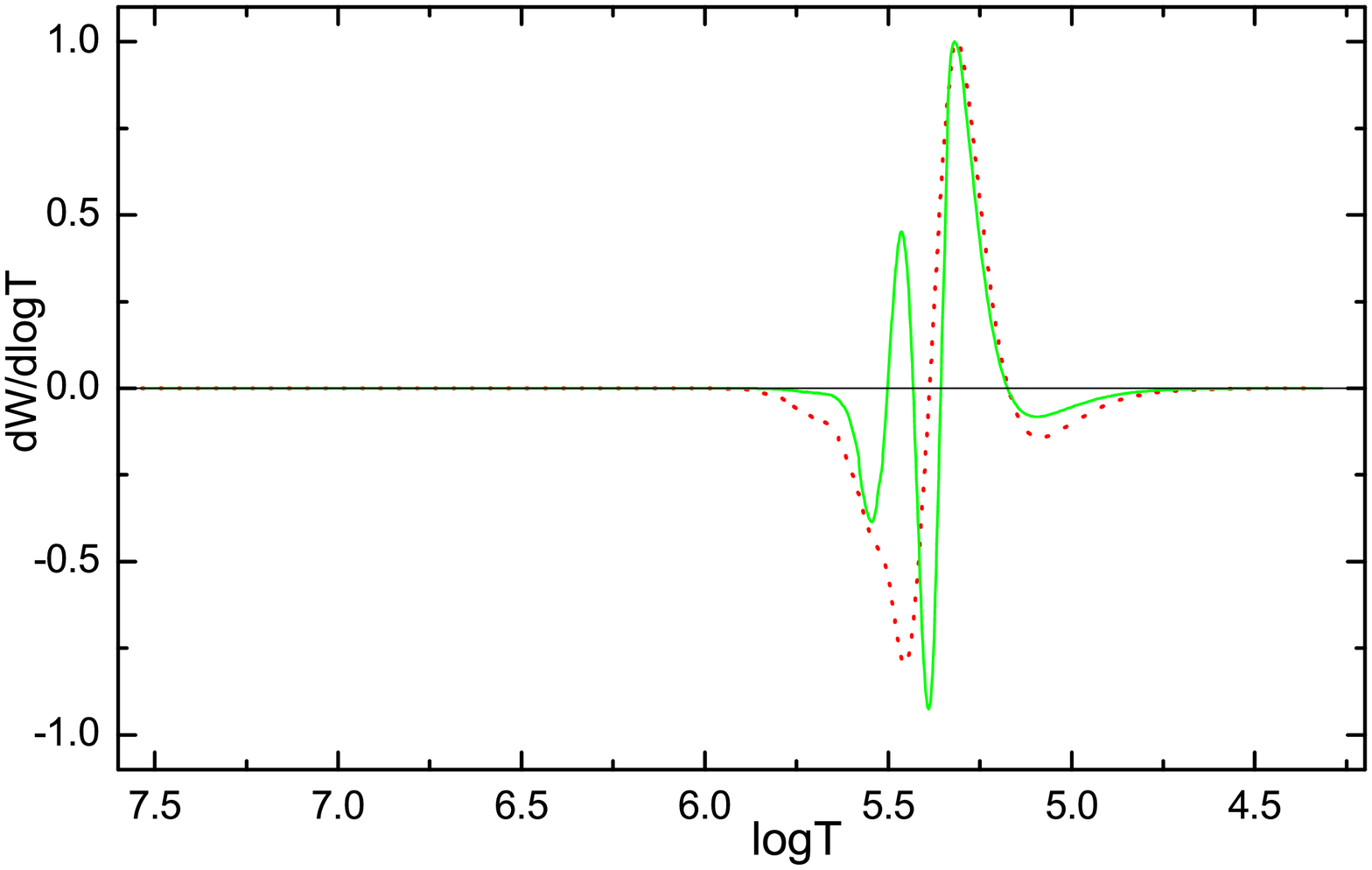}
 \includegraphics[clip,width=\columnwidth,height=4.8cm]{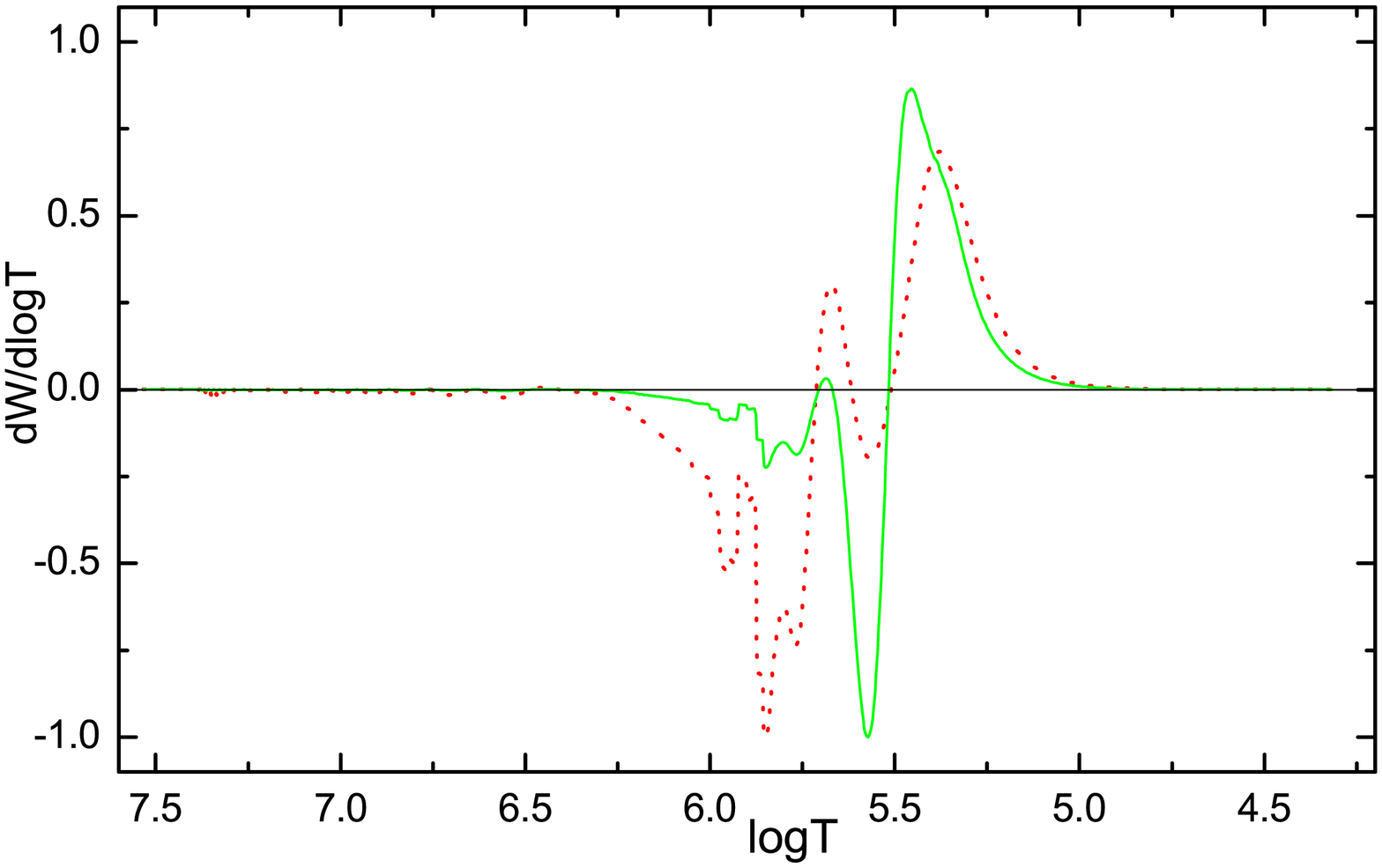}
 \includegraphics[clip,width=\columnwidth]{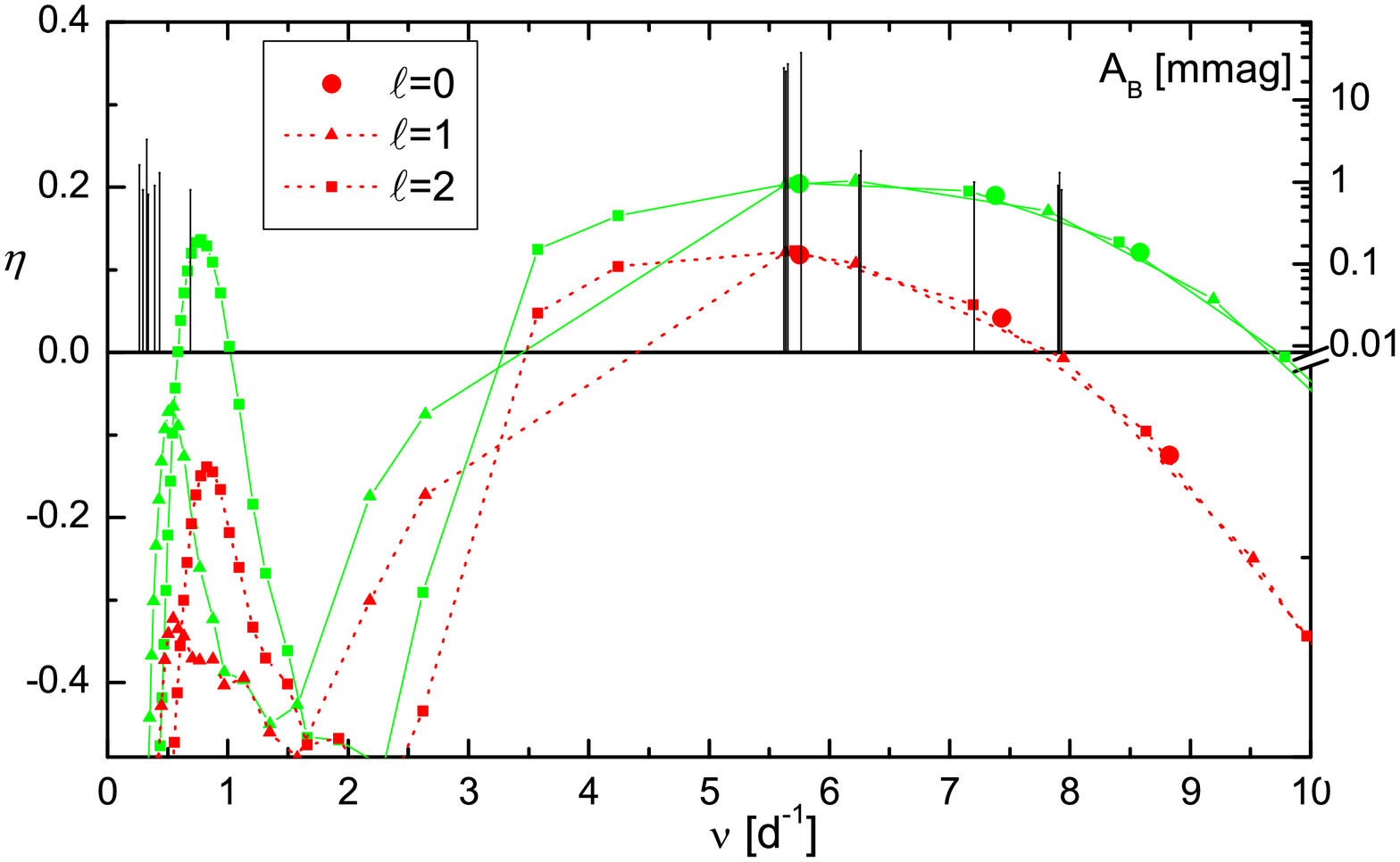}
   \caption{The upper panel: the run of the standard (dotted line) and modified (solid line) Rosseland mean OPLIB opacity
(the left hand Y-axis) and its temperature derivative (the right hand Y-axis) for the model: $M=9.5M_{\odot}$, $Z=0.015$,
$\alpha_{\rm{ov}}=0.07$, $\log{T_{\rm eff}}=4.3399$, $\log{L/L_{\odot}}=3.917$. The $\kappa$ profile was modified by adding
opacity at $\log T_0=5.3$ and 5.46 and the parameters of this modification are given in the legend. The middle panels:
a comparison of the differential work integral for the radial second overtone mode, $\ell=0,~p_3$, and high overtone
quadrupole g mode, $\ell=2,~g_{16}$, in models computed with the standard and modified opacity profile. The bottom panel:
a comparison of the corresponding normalized instability parameter, $\eta$.}
\label{fig3}
\end{figure}

\begin{figure}
 \includegraphics[clip,width=\columnwidth]{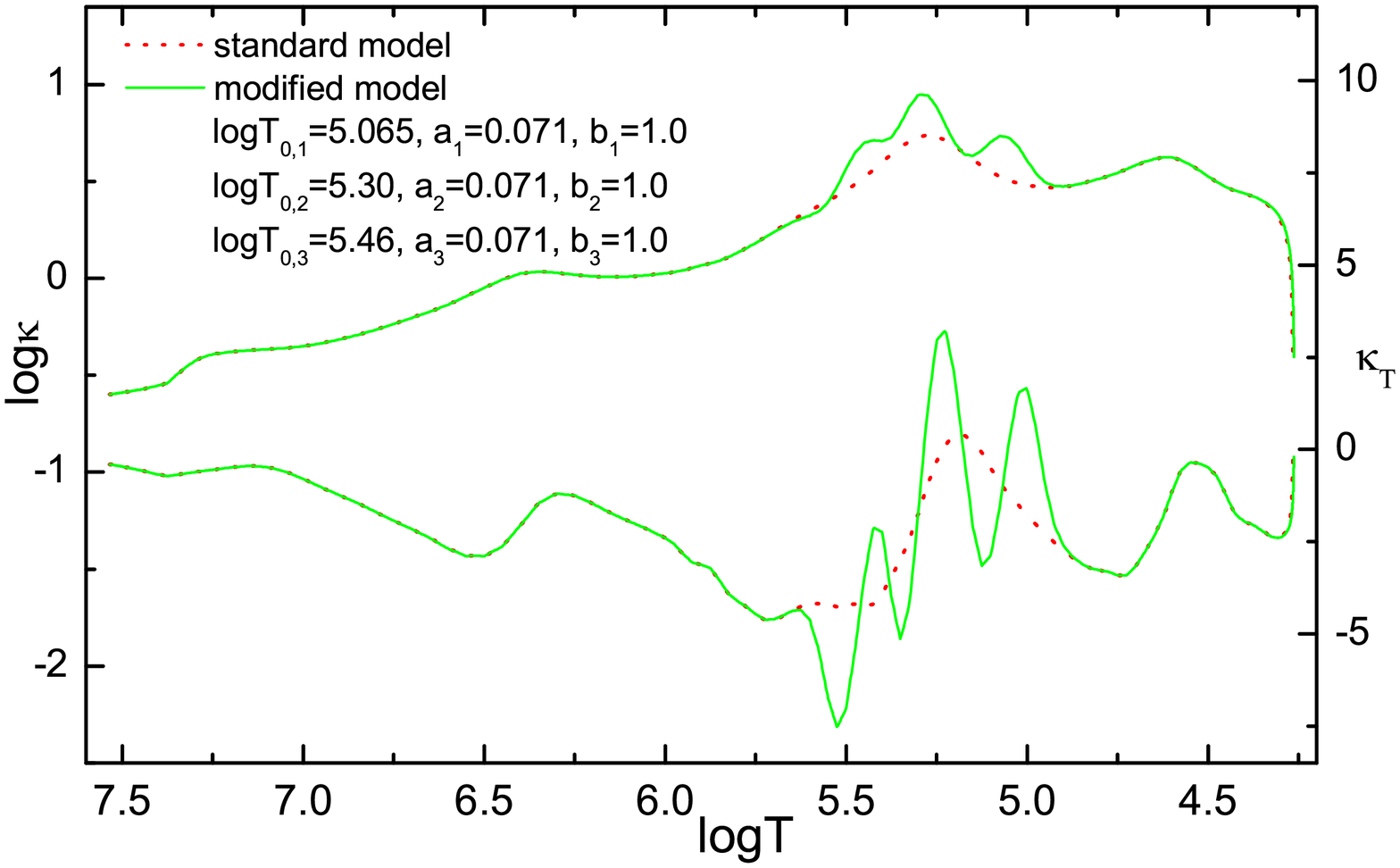}
 \includegraphics[clip,width=\columnwidth]{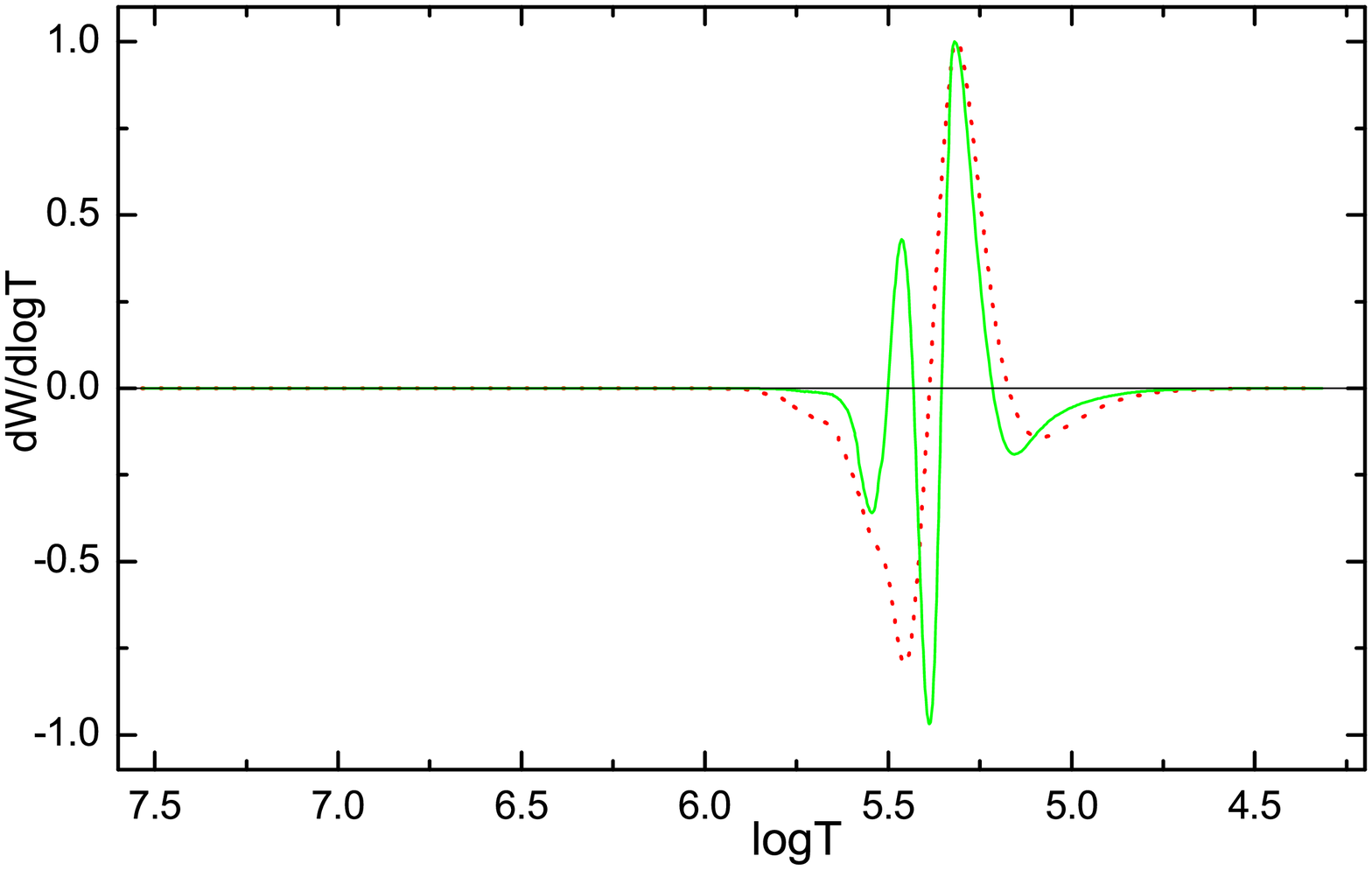}
 \includegraphics[clip,width=\columnwidth]{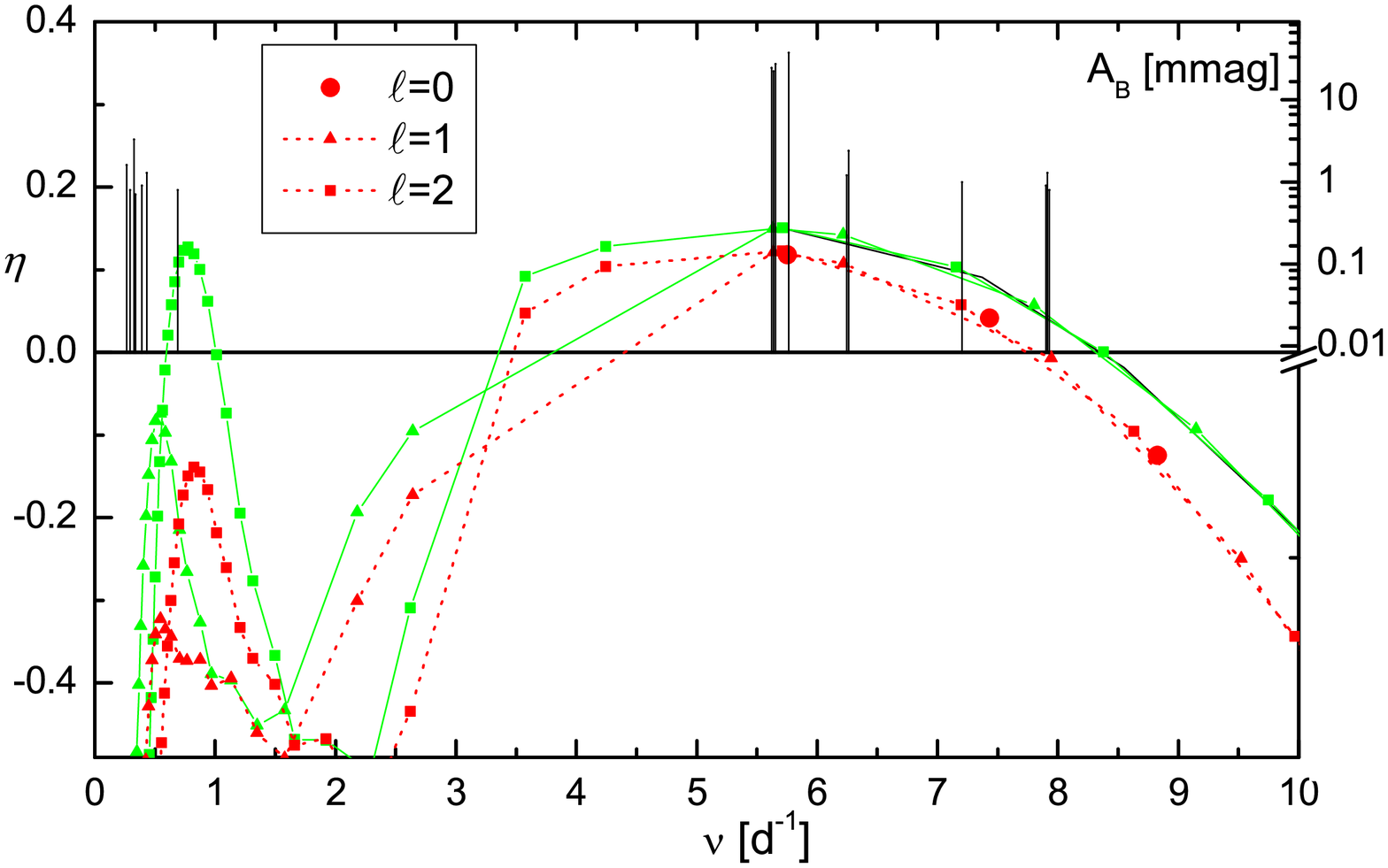}
   \caption{A similar figure as Fig.\,3 but the $\kappa$ profile was additionally modified at $\log T_0=5.065$. The differential
work integral for the mode $\ell=2,~g_{16}$ is not shown because it looks very similar to that shown in Fig.\,3.  }
\label{fig4}
\end{figure}

Then, to reproduce the observed frequency ranges as well as the values of some frequency peaks, we determined the corrections
to $\kappa(T)$ with the following steps $\Delta a=0.001$, $\Delta b=0.05$ and $\Delta T_0=0.005$ in the range $\log T\in (5.0 - 5.5)$.
For each opacity database, we found many models that reproduce these observed features. In Fig.\,5, we give examples
of such models for the three used opacity data sets: OPLIB (the top panel), OPAL (the middle panel) and OP (the bottom panel).
The parameters of the models are given in Table\,1 and modifications of the opacity profile are as follows:
\begin{itemize}
\item \textbf{OPLIB}: $\log T_{0,1}=5.06,~a_1=0.071,~b_1=0.3;\\~~~~~~~~~~~~~~~~~~~~~~~\log T_{0,2}=5.46,~a_2=0.224,~b_2=1.5,$
\item \textbf{OPAL}:  $\log T_{0,1}=5.30,~a_1=0.082,~b_1=0.5;\\~~~~~~~~~~~~~~~~~~~~~~\log T_{0,2}=5.46,~a_2=0.082,~b_2=1.5,$
\item \textbf{OP}:    $\log T_{0,1}=5.20,~a_1=0.071,~b_1=0.5;\\~~~~~~~~~~~~~~~~\log T_{0,2}=5.46,~a_2=0.071,~b_2=1.0.$
\end{itemize}
\begin{figure}
 \includegraphics[clip,width=\columnwidth]{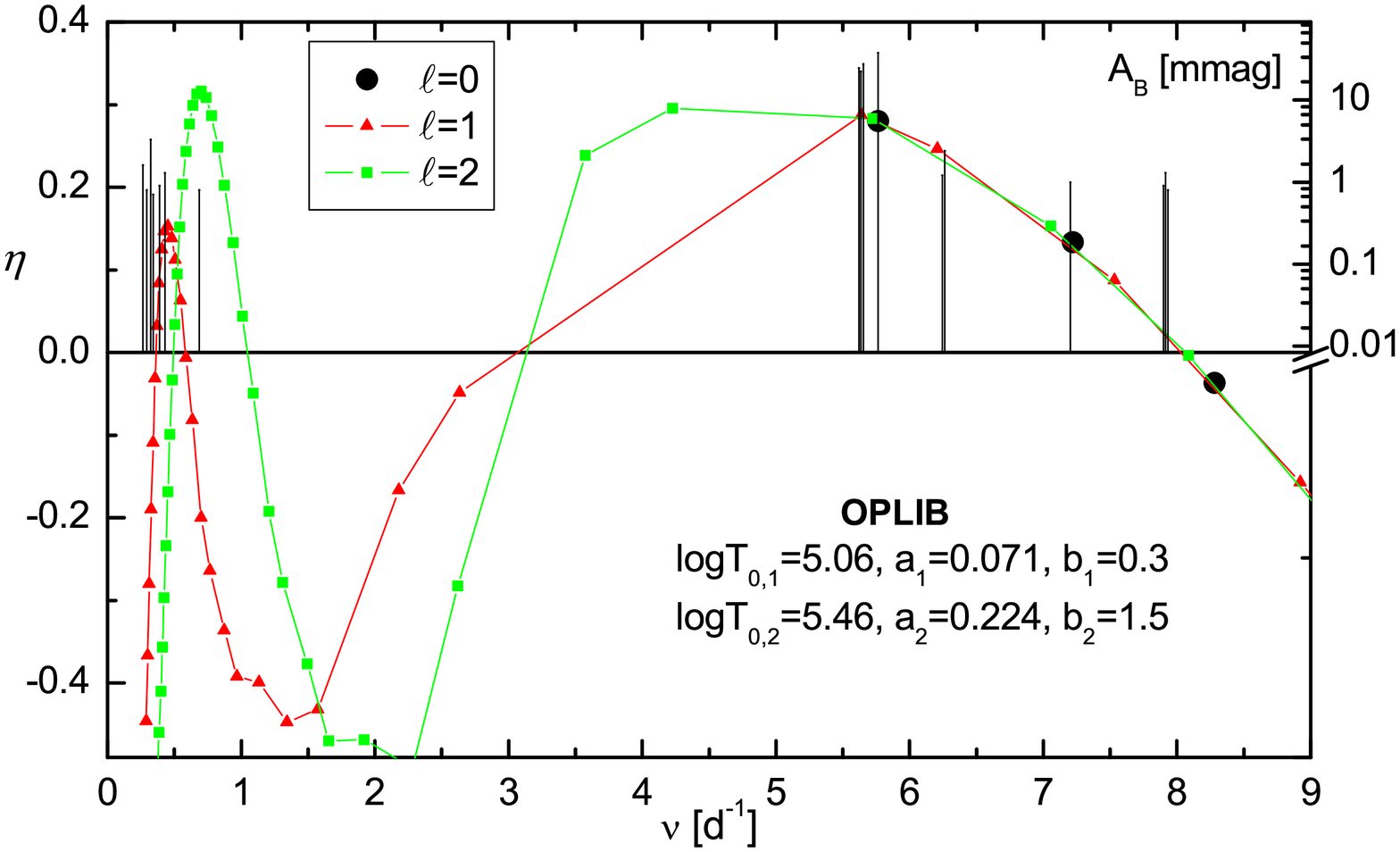}
 \includegraphics[clip,width=\columnwidth]{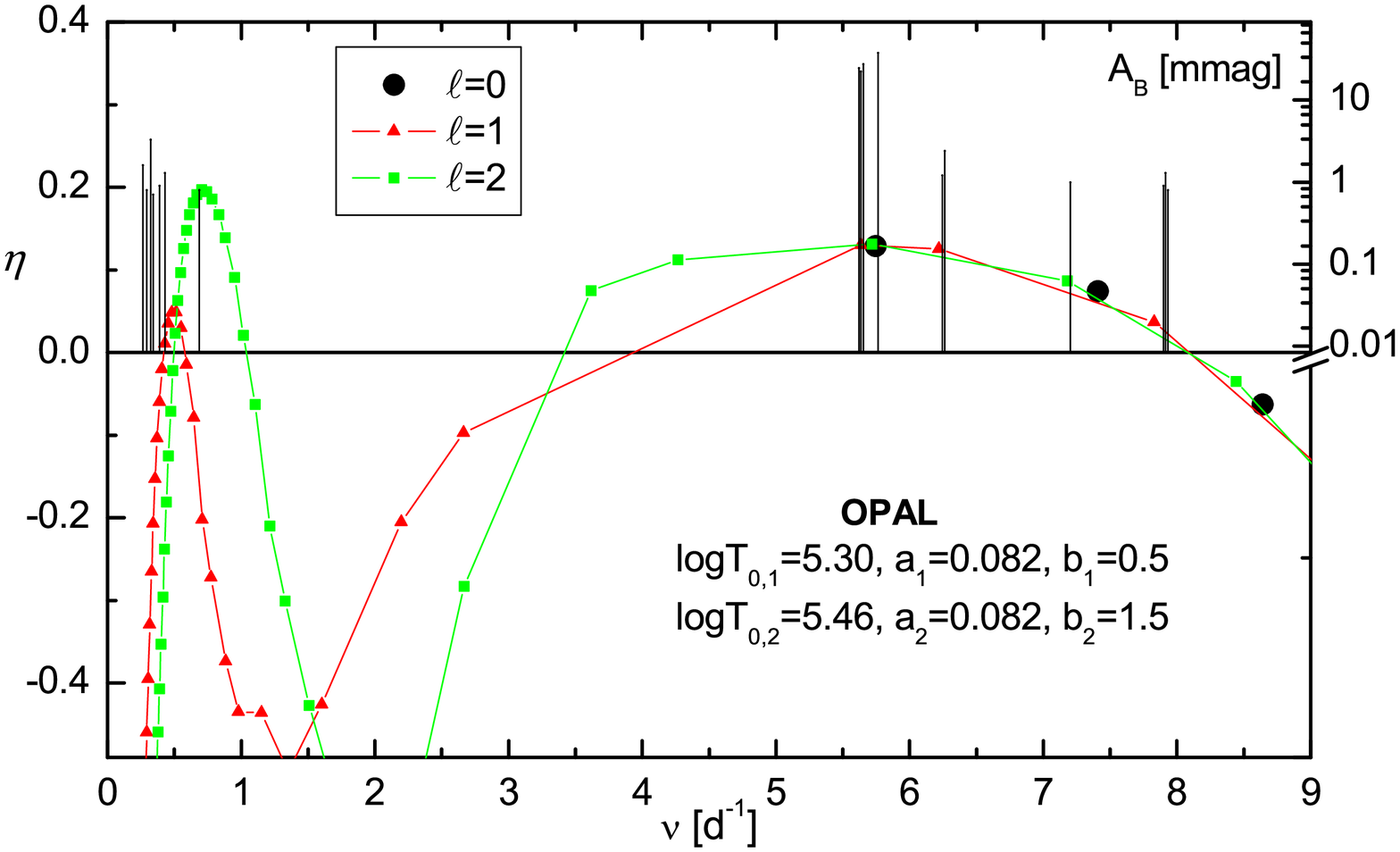}
 \includegraphics[clip,width=\columnwidth]{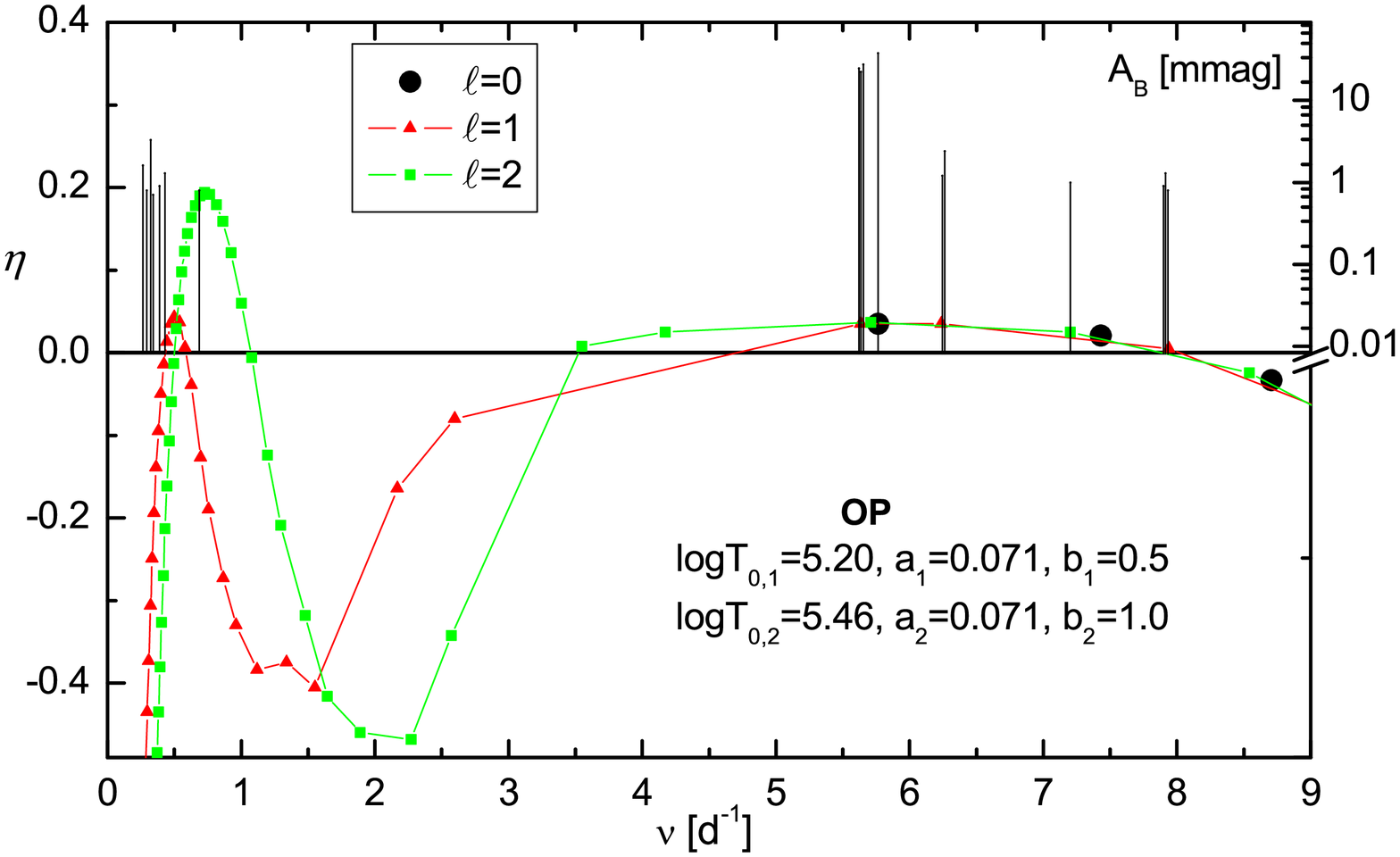}
   \caption{The normalized instability parameter, $\eta$, as a function of mode frequency for models that roughly
reproduce some observed frequencies and the range of instability for both p and g modes of $\nu$ Eri. For each type of opacity data
(OPLIB, OPAL, OP), there is shown one model of this kind. The OPLIB model:
$M=9.5M_\odot$, $\log T_{\rm eff}=4.3367,~\log L/L_\odot=3.916,~Z=0.015, ~\alpha_{\rm ov}=0.07$;
the OPAL model: $M=9.2~M_\odot, ~\log T_{\rm eff}=4.3348,~\log L/L_\odot=3.886,~Z=0.015,~\alpha_{\rm ov}=0.15$,
the OP model: $M=9.6M_\odot,~\log T_{\rm eff}=4.3358,~\log L/L_\odot=3.890,~Z=0.0185,~\alpha_{\rm ov}=0.0$.
The initial value of hydrogen was $X_0=0.7$ in each case. The parameters ($T_0,~a,~b$) of the modified opacity profile are
given in the legend.}
\label{fig5}
\end{figure}

As one can see, all of these models have unstable modes in the high frequency range (5-8) d$^{-1}$ which perfectly agrees
with observations.
Moreover, the theoretical frequencies of the highest amplitude p modes agree with the observed values to the second decimal place, in line
with mode identification (e.g., Daszy\'nska-Daszkiewicz et al. 2005, Daszy\'nska-Daszkiewicz \& Walczak 2010.)
The instability range of high order g modes is worse reproduced and all modes with frequencies below 0.35 d$^{-1}$
do not have their theoretical counterparts.
There are only a few unstable dipole modes. Quadrupole modes are more unstable but they are shifted to higher frequencies.
This deficiency can be partially explained by rotational splitting, in particular, if the core was really spinning faster
as suggested by Pamyatnykh, Handler \& Dziembowski (2004).

\section{How to constrain the modifications of stellar opacities?}

As we have mentioned at the end of the previous section, it is possible to find many models that can account simultaneously
for the observed frequency ranges and roughly fit the values of some frequencies. Thus, the question is:
which modification of the opacity profile is more likely?
First, one can think about drawing some conclusions from chemical compositions determined from high-resolution spectra.
Such analysis by Morel et al (2006) and, more recently, by Nieva \& Przybilla (2012) points to normal (solar-like) abundances
for $\nu$ Eri. However, these photospheric abundances do not necessarily reflect the subphotospheric values. Moreover,
the values of opacities do not depend only on the abundance of individual elements but also on how they are computed
(e.g., on the number of fine structure energy levels per ion taken into account).

Thus, we need an observable that would be sensitive to the opacity of the subphotospheric layers near the Z-bump.
Such an indicator is the relative amplitude of radiative flux perturbation at the level of the photosphere, which is called
the parameter $f$.
The value of $f$ is complex and results from linear non-adiabatic computations of stellar pulsations.
This asteroseismic probe was introduced by Daszy\'nska-Daszkiewicz, Dziembowski \& Pamyatnykh (2003) and in the case of
B-type pulsators, it is strongly sensitive to the metallicity and opacity data (Daszy\'nska-Daszkiewicz,
Dziembowski \& Pamyatnykh 2005). The empirical counterparts of $f$ are derived from multicolour photometry and
radial velocity data by means of the method proposed by Daszy\'nska-Daszkiewicz, Dziembowski \& Pamyatnykh (2003).
An interesting result from the last studies of the parameter $f$ for $\nu$ Eri is a preference of the OPAL tables
by p modes and a better agreement with the OP opacity models for high order g modes (Daszy\'nska-Daszkiewicz \& Walczak 2010).
Here, we will make use of the empirical values of $f$ for the radial fundamental mode $\nu=5.76324$ d$^{-1}$ derived
from the Str\"omgren amplitudes and phases and radial velocity variations from the last multi-site campaigns.
It should be mentioned that the empirical values of $f$ weakly depend on the stellar parameters in the allowed ranges
whereas they do depend on model atmospheres.
In Table\,1, we give the parameters of the seismic models found in the previous section and depicted in Fig.\,5.
There are also provided the theoretical and empirical values of the absolute value of the $f$ parameter, $|f|$,
and the phase lag, $\Psi={\rm arg}(f)-180^\circ$, for the radial fundamental mode of $\nu$ Eri.
The value of $\Psi$ gives the phase shift between the maximum temperature and the minimum radius.
The corresponding empirical values of $|f|$ and $\Psi$ were derived by adopting
both LTE models (Kurucz 2004) and NLTE models (Lanz \& Hubeny 2007) atmospheres.

\begin{table*}
\begin{center}
\caption{Parameters of the seismic models of $\nu$ Eri with the modified opacity profile, shown in Fig.\,5 and Fig.\,6 (bestOPLIB).
Columns from left to right are: opacity data, mass, $M/M_\odot$, effective temperature, $\log T_{\rm{eff}}$, luminosity,
$\log{L/L_{\odot}}$, metallicity, $Z$, overshooting parameter, $\alpha_{\rm{ov}}$, the theoretical values of $(|f|,~\Psi)$ for
the radial fundamental mode and their empirical counterparts. The latter were computed with both the LTE and NLTE atmosphere models.}
\begin{tabular}{|l|c|c|c|c|c|c|c|c|}
\hline
model & $M$ [$M_\odot$] & $\log{T_{\rm eff}}$ & $\log{L/L_\odot}$ & $Z$ & $\alpha_{\rm{ov}}$
& $(|f|,~\Psi)_{\rm teo}$ & $(|f|,~\Psi)_{\rm emp}^{\rm LTE}$ & $(|f|,~\Psi)_{\rm emp}^{\rm NLTE}$\\
\hline
OPLIB &9.5  &4.337  &3.92  &0.015  &0.07  & $(5.33,~-20.46^\circ)$  & $(9.42(21),~-4.86^{\circ}(1.26))$ & $(9.01(26),~-4.88^{\circ}(1.66))$ \\
\hline
OPAL & 9.2 & 4.335 &  3.89 & 0.015 & 0.15 & $(6.52,~-15.70^\circ)$  & $(9.31(24),~-4.83^{\circ}(1.46))$  & $(8.91(28),~-4.84^{\circ}(1.79))$ \\
\hline
OP   &  9.6 & 4.336 & 3.89 & 0.0185 & 0.0  &  $(8.44,~-0.15^{\circ})$  & $(9.27(23),~-4.84^{\circ}(1.44))$  & $(8.86(28),~-4.85^{\circ}(1.81))$ \\
\hline
\hline
bestOPLIB & 9.0 & 4.331 & 3.86 & 0.015 & 0.163 & $(8.65,~-2.23^{\circ})$  & $(9.20(28),~-4.82^{\circ}(1.75))$  &  $(8.82(31),~-4.82^{\circ}(1.98))$ \\
\hline
\end{tabular}
\end{center}
\end{table*}

As one can see, there are significant differences in the values of $(|f|,~\Psi)$ between models computed with different opacity data as well
as a large disagreement with their empirical counterparts.
Thus now the aim is to find the $\kappa-$modified models that will reproduce additionally the empirical value of $f$ of
the dominant mode.
It appeared that fitting the parameter $f$ significantly reduces the number of seismic models. In fact, it is quite hard
to find the models that simultaneously reproduce the observed frequencies, the range of instability for both p and g modes
and the empirical value of the nonadiabatic parameter $f$.
\begin{figure}
 \includegraphics[clip,width=\columnwidth]{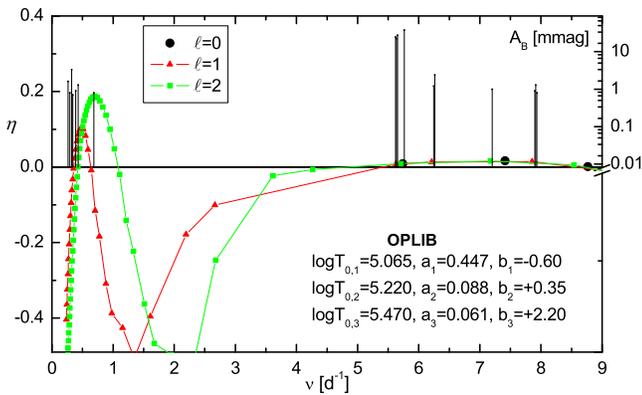}
   \caption{The OPLIB seismic model, which reproduces the observed frequencies of p modes, the range of instability for
both p and g modes, and the empirical value of the nonadiabatic parameter $f$ for the dominant frequency corresponding
to the radial fundamental mode. The parameters of the model are: $M=9.0~M_\odot$, $\log T_{\rm eff}=4.3314$, $Z=0.015$, $X_0=0.7$,
$\alpha_{\rm ov}=0.163$. The values of temperatures and parameters $(a,~b)$ at which the $\kappa$ profile was modified are given
in the legend. }
\label{fig6}
\end{figure}
\begin{figure}
 \includegraphics[clip,width=\columnwidth]{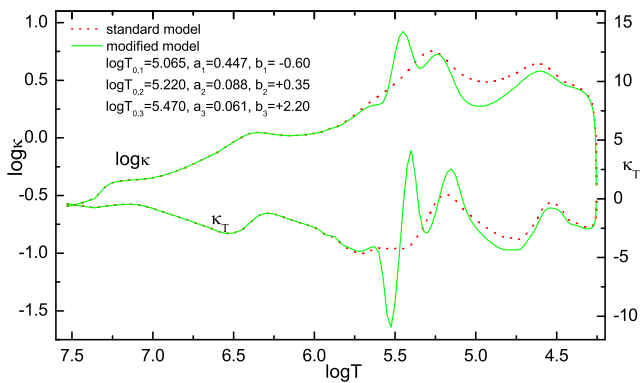}
   \caption{The run of the modified Rosseland mean opacity (the left-hand Y-axis) and its temperature derivative
(the right-hand Y-axis) for the model presented in Fig.\,6.}
\label{fig7}
\end{figure}

Nevertheless, we succeeded in finding such a model with the OPLIB data and in Fig.\,6 we depict its instability parameter
$\eta$ as a function of the frequency. The model has the following parameters: ~$M= 9.0~M_\odot$, $\log T_{\rm eff}=4.3314$,
$\log L/L_\odot= 3.859$, $Z=0.015$, $\alpha_{\rm ov}=0.163$, and the modification of the opacity profile is:

\begin{itemize}
\item \textbf{bestOPLIB}: $\log T_{0,1}=5.065,~a_1=0.447,~b_1=-0.60;\\~~~~~~~~~~~~~~~~~~~~~~~\log T_{0,2}=5.22,~a_1=0.088,~b_1=0.35;\\~~~~~~~~~~~~~~~~~~~~~~~\log T_{0,3}=5.47,~a_2=0.061,~b_1=2.20$.
\end{itemize}

Our best model has a greatly reduced value of the mean opacity near the temperature $\log{T}=5.065$. This was necessary to fit
the empirical values of the parameter $f$.
On the other hand, the mean opacity has to be slightly increased near $\log{T}=5.22$ and significantly increased near
$\log{T}=5.47$ in order to fulfill the instability requirement for p and g modes, respectively.
The modified $\kappa$ profile together with its derivative with respect to temperature, $\kappa_T$, is shown in Fig.\,7.
The consequences of this $\kappa$ modification on the flux eigenfunction $f(\log T)$ are large as can be judged from Fig.\,8,
where the absolute value, $|f|$, and the phase lag, $\Psi={\rm arg}(f)-180^\circ$, are plotted.
The photospheric values of the empirical $(|f|,~\Psi)$ are marked with the short horizontal lines.
Both, the absolute value, $|f|$, and the phase, $\Psi$, are significantly modified around the minimum values of the the opacity
derivative, $\kappa_T$, with the proviso that $|f|$ reaches the maximum values at $\log T\approx 5.53$ where the derivative $\kappa_T$ reaches the minimum,
whereas the value of $\Psi$ is the maximum at $\log T\approx 5.41$ where $\kappa_T$ obtains the maximum.
It is worth mentioning that many models that reproduce the empirical parameter $f$ for the dominant frequency corresponding
to the radial fundamental mode could not account for the instability of high order g modes. This is because to excite the g modes
one has to increase significantly the opacity which, in turn, results in increasing the theoretical value of the phase lag $\Psi$.

Despite such large modification of the opacity profile, the value coming out of the photosphere is not much changed
and the parameter $f$ is mostly adjusted to the empirical value by reducing the opacity at $\log T_{0,1}=5.065$ which
does not affect the instability of g modes.
The theoretical values $|f|$ and $\Psi$ of the model from  Fig.\,6 and the corresponding empirical values are given in the last line of Table\,1.
As one can see, the theoretical and empirical values of $|f|$ are in excellent agreement if the NLTE model atmospheres are used.
With the LTE models, the agreement is achieved if the $2\sigma$ error is allowed.
For the argument, $\Psi$, the compatibility is within the $3\sigma$ error regardless of which model atmospheres are used.
\begin{figure}
 \includegraphics[clip,width=\columnwidth]{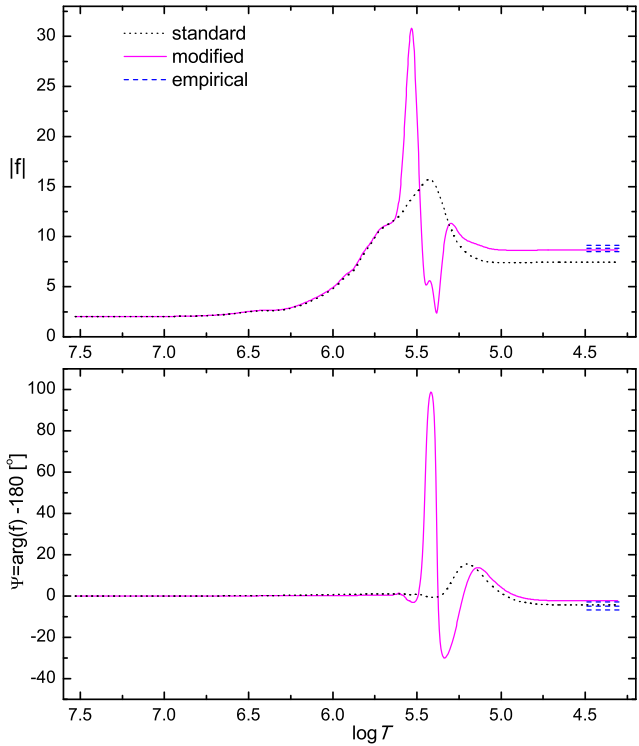}
   \caption{The absolute value (the upper panel) and the phase lag (the bottom panel) of the flux eigenfunction $f(\log T)$
for the model presented in Fig.\,6.  The dashed blue horizontal line corresponds
to the empirical values derived with the NLTE atmospheres.}
\label{fig8}
\end{figure}

\section{Additional effects of the opacity modification}

\subsection{The effect of convection in the Z-bump}

Despite the general considerations that convective transport is negligible in stellar models with masses corresponding to
$\beta$ Cep variable ($M\approx 7-16~M_\odot$), it can be important at the depth where the local maxima of the opacity occur.
As was shown by Cantiello et al. (2009), the efficiency of the Z-bump convection increases with increasing metallicity,
decreasing effective temperature
and increasing total luminosity. These three facts were confirmed by the observational data.
Moreover, fast rotation can enhance this subsurface convection (Maeder et al. 2008).

In the previous section we showed that the nonadiabatic parameter $f$ is a powerful probe of modification of the stellar
opacities in the depth range $\log T=5.0 - 5.5$. Because of this important diagnostic property, here we will check the
sensitivity of the flux eigenfunction $f(\log T)$ to the efficiency of convection in the Z-bump layer.
The convective efficiency is measured by the value of the mixing length parameter, $\alpha_{\rm MLT}$.

We consider the modified OP opacity profile with the parameters:
$\log T_{0,1}=5.15,~a_1=0.071,~b_1=1.0;~~\log T_{0,2}=5.46,~a_1=0.071,~b_1=1.0$,
and the two values of the MLT parameter: $\alpha_{\rm MLT}=0.5$ and 5.0. This second large value of $\alpha_{\rm MLT}$ is chosen,
firstly,
to show the effect and, secondly, because such values of $\alpha_{\rm MLT}$ were considered at some depth of the solar convective zone.
The depth-dependence of $\alpha_{\rm MLT}$ was studied by, e.g., Schlattl et al. (1997) and more recently by Magic, Weiss \& Asplund (2015).
\begin{figure}
 \includegraphics[clip,width=\columnwidth]{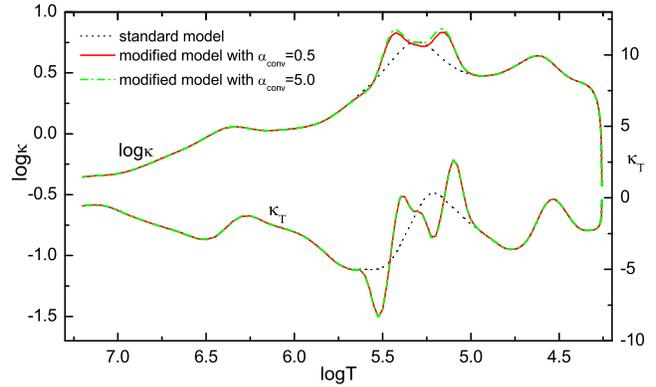}
   \caption{The run of the standard (dotted line) and modified (solid line) profiles of the Rosseland mean OP opacity
(the left hand Y-axis) for the two values of the MLT parameter, $\alpha_{\rm MLT}=0.5$ and $\alpha_{\rm MLT}=5.0$.
The corresponding values of the temperature derivatives of $\kappa$ are given on the right hand Y-axis.
The parameter of the model are: $Z=0.0185$, $M=9.63~M_\odot$, $\log T_{\rm eff}\approx 4.334$, $\log L/L_\odot\approx 3.89$.}
\label{fig8}
\end{figure}

In Fig.\,9, we plot the standard and modified profiles of the Rosseland mean OP opacity (the left-hand Y-axis)
for the two values of the MLT parameter, $\alpha_{\rm MLT}=0.5$ and $\alpha_{\rm MLT}=5.0$. The corresponding
$\kappa$ derivatives over temperature are depicted as well (the right-hand Y-axis). The model parameters are: $Z=0.0185$,
$M=9.63~M_\odot$, $\log T_{\rm eff}\approx 4.334$, $\log L/L_\odot\approx 3.89$. As one can see, the change of $\alpha_{\rm MLT}$
is only revealed near the local maxima of $\kappa$.
For the model with the modified opacity and $\alpha_{\rm MLT}=5.0$, the efficiency of convection reaches about 45\% in the vicinity
of the Z-bump layer.
The effect of $\alpha_{\rm MLT}$ on the differential work integral is shown in Fig.\,10 where its value is depicted for
the radial fundamental mode, corresponding to $\nu=5.7625$ d$^{-1}$, and the high-overtone g mode, $\ell=1, g_{14}$,
corresponding to $\nu=0.5003$ d$^{-1}$.
Although the effect of $\alpha_{\rm MLT}$ seems to be negligible, the result is that in the model with $\alpha_{\rm MLT}=0.5$
the radial fundamental mode is stable whereas in the model with $\alpha_{\rm MLT}=5.0$  it is unstable because of smaller
damping at $\log T\approx 5.35$ and $\log T\approx 5.55$.
In the case of the g mode, the damping around $\log T=5.55$ is slightly reduced in the model with $\alpha_{\rm MLT}=5.0$
and the instability parameter, $\eta$, is a little larger.
\begin{figure}
 \includegraphics[clip,width=\columnwidth]{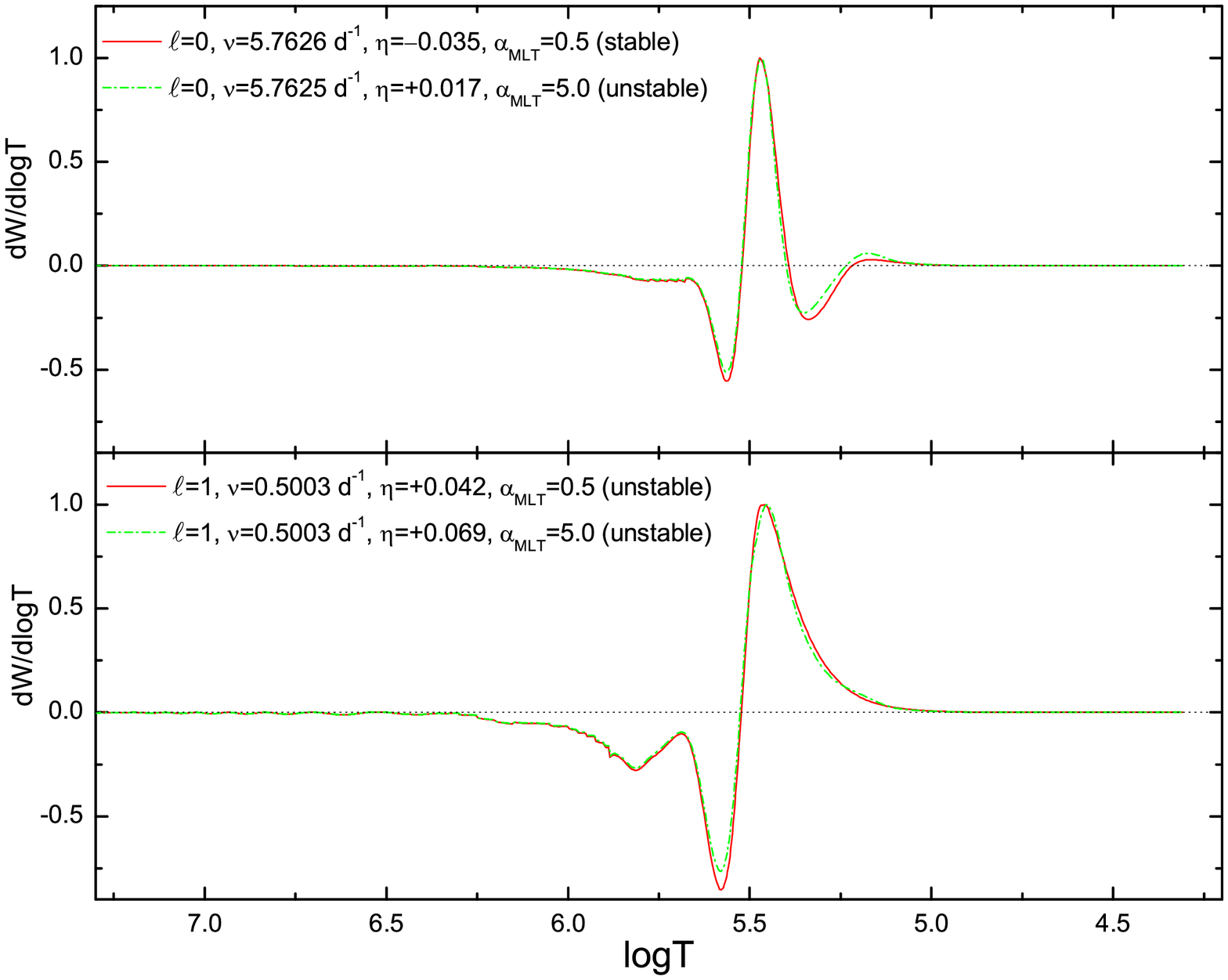}
   \caption{The differential work integral for the radial fundamental mode (the top panel) and dipole gravity mode (the bottom panel)
for the models with the modified $\kappa$ profile (cf. Fig.\,9). The effect of the MLT parameter is illustrated.
For $\alpha_{\rm MLT}=0.5$ the $\ell=0,p_1$ mode is stable whereas for $\alpha_{\rm MLT}=5.0$ it is excited.}
\label{fig10}
\end{figure}
\begin{figure}
 \includegraphics[clip,width=\columnwidth]{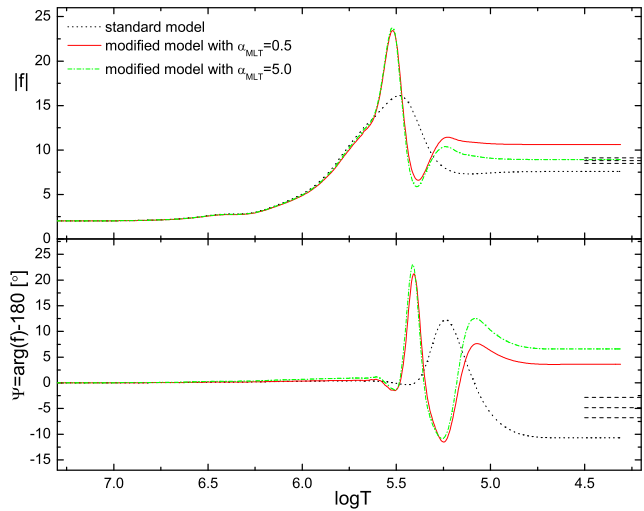}
   \caption{The effect of the MLT parameter on the flux eigenfunction $f(\log T)$. The upper and bottom panels show the absolute
value and the phase of $f(\log T)$, respectively. The dotted line corresponds to $f(\log T)$ in the model with
the standard OP opacities. The short horizontal lines correspond to the empirical values derived with the NLTE atmospheres.}
\label{fig11}
\end{figure}

How the value of $\alpha_{\rm MLT}$ affects the flux eigenfunction, $f(\log T)$, is presented in Fig.\,11.
The dashed and solid lines correspond to the opacity-modified models with $\alpha_{\rm MLT}=0.5$ and 5.0, respectively.
The model with the standard opacity profile is plotted with the dotted line. As one can see,
the dependence of $f(\log T)$ on the MLT parameter is important and the difference is larger than the observational error
of the parameter $f$.
The adjustment of the value of $\alpha_{\rm MLT}$ is beyond the scope of this paper but one has to keep in mind this fact
when studying the properties of the Z-bump in the $\beta$ Cep star models.

\subsection{Avoided crossing of radial modes}

When studying models with the modified $\kappa$ profile, we encountered the phenomenon occurring so far only for nonradial modes.
It appears that in models with a significant increase of opacity at certain temperatures, the radial modes
can experience the avoided-crossing phenomenon, which in standard massive main sequence models occurs only for nonradial oscillations.
Till now the avoided crossing for radial modes was noticed in models of neutron stars (Gondek, Haensel \& Zdunik 1997,
Gondek \& Zdunik 1999) and Cepheids (Buchler, Yecko \& Kollath 1997).
\begin{figure}
 \includegraphics[clip,width=\columnwidth]{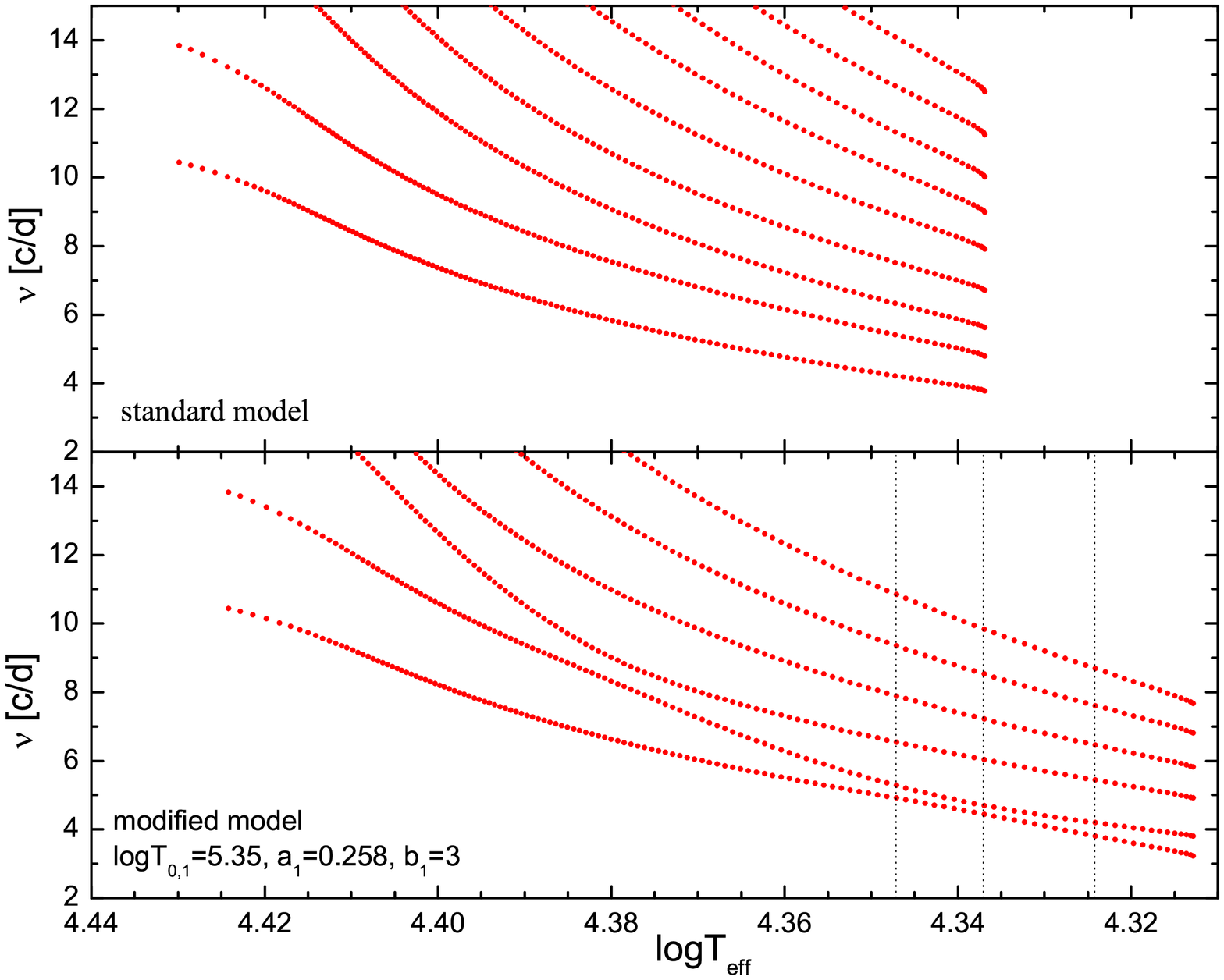}
   \caption{Evolution of eigenfrequencies of radial modes in a model with $M=11.3 M_{\sun}$ from ZAMS to TAMS, computed with the standard OP opacities
(the upper panel) and modified ones (the bottom panel) for the metallicity $Z=0.0185$. In the later case, the opacity was increased by a factor of 4
at the depth $\log T=5.35$. Dotted vertical lines mark positions of the selected models before, during and after the avoided
crossing of the fundamental mode and first overtone.}
\label{fig12}
\end{figure}

In certain cases, an increase of the local opacity produces the two resonance cavities
for radial modes and low order overtones can exchange their dynamic behaviour in stellar interiors.
An example is shown in Fig.\,12, where the evolution of eigenfrequencies of the six radial overtones for models with $M=11.3 M_{\sun}$ is shown.
The top and bottom panels correspond to the standard and modified models, respectively.
To emphasize the effect, in the modified model the OP opacities were increased by a factor of 4 around the depth $\log T=5.35$.
As one can see, at certain evolutionary stages the first and second overtone radial modes  $(\log T_{\rm eff}\approx 4.38)$
or the fundamental and first overtone radial modes $(\log T_{\rm eff}\approx 4.34)$ can get close to each other.

The standard and modified $\kappa$ profiles are depicted in the top panel of Fig.\,13. Note that maximum difference in the opacity
between two models is a factor of about 2.3, not 4, and the position of this maximum is at $\log T\approx 5.25$, not 5.35,
as in the standard and modified opacity tables. This is due to the fact that two evolutionary models differ one from another
by distribution of both temperature and density in the interiors. Therefore, the opacity coefficient is interpolated along
different $\log T$ - $\log \rho$ lines in the tables.

During the avoided crossing of radial modes, overtones exchange their oscillatory properties, in particular the distribution
of the kinetic energy inside a model. This can be seen from the lower panels of Fig.\,13
where we show the run of the kinetic energy density as a function of the depth, $\log T$, for the three lowest radial modes,
$\ell=0, ~p_1,~p_2,~p_3$, at three close evolutionary stages: before, during and after the avoided crossing of the fundamental mode
and first overtone. These three selected models are marked in Fig.\,12 by vertical lines.
In this case after the avoided crossing the fundamental and first overtone modes exchange their dynamical behaviour in the interior:
the distribution of the kinetic energy of fundamental mode is now very similar to that of the first overtone before
the avoided crossing, and the distribution of the kinetic energy of the first overtone is very similar to that of the fundamental mode
before the avoided crossing. It should be mentioned that during the avoided crossing both modes have the same normalized kinetic energy.
At the closest approach the kinetic energy is nearly equally divided between the regions above and below $\log T\approx 5.7$,
which is the lower border of the potential barrier formed by the opacity increase around $\log T=5.35$.
\begin{figure}
 \includegraphics[clip,width=\columnwidth]{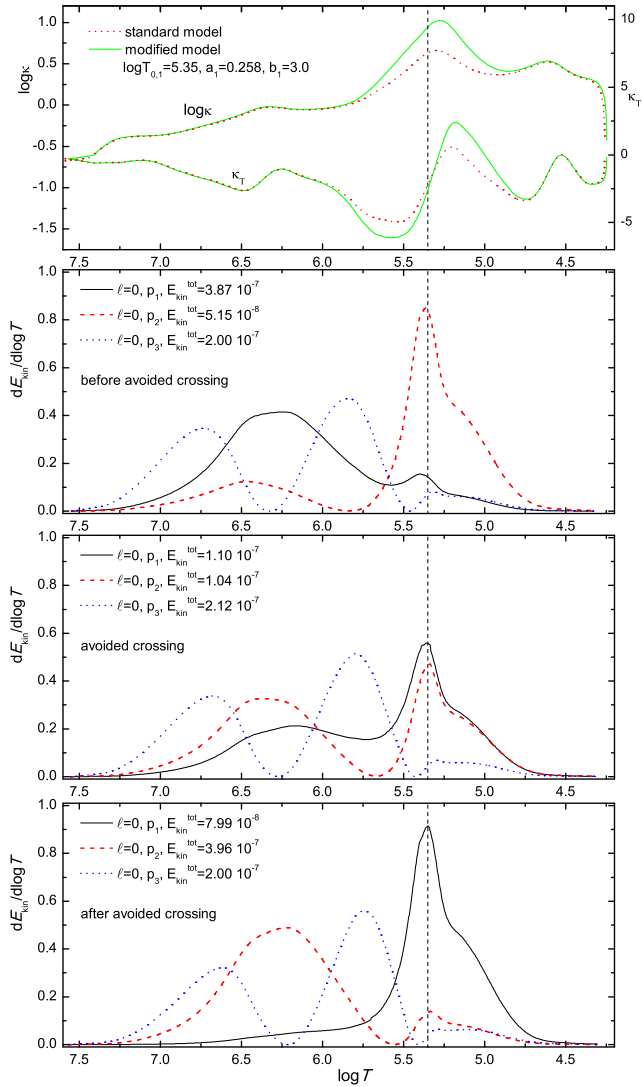}
   \caption{The top panel: the run of the opacity and its temperature derivative in the models computed with the standard
and modified $\kappa$ as described in the text. The vertical line corresponds to $\log T=5.35$ at which the opacity was increased
by 4. The panels below show the kinetic energy density of the three radial modes in the $\kappa$ modified models before, at
and after the avoided crossing. These models are marked by vertical lines in Fig.\,12.}
\label{fig13}
\end{figure}

\section{Conclusions and future works} \label{conclusions}

Seismic modelling of pulsating stars that exhibit both p modes and higher order g modes is still a challenging task.
In this paper, we aimed at finding complex seismic models of the well-known pulsator $\nu$ Eri which reproduce simultaneously
the observed frequency range, the values of some individual frequencies and the empirical value of the nonadiabatic parameter $f$
for the dominant radial mode.

We based our computations on the three sources of the most widely used opacity data, OPLIB, OPAL and OP,
and found that in each case only models with the modified profile of the Rosseland mean opacity can account for the instability of the observed modes.
A very important result is that adding the requirement of fitting the nonadiabatic parameter $f$
greatly reduces the number of $\kappa$ modifications and, in fact, the good news is that it is very difficult to pick out such a model.
We found that the model computed with the OPLIB data modified at the three depths, $\log T=5.06,~5.22,~5.47$,
best meets all the above conditions. Thus, there is some indication that these opacities are preferred.
A large increase (more than three times) of the opacity at $\log T=5.47$ was indispensable to get the instability of g modes
whereas a reduction of the opacity by 65\% at $\log T=5.06$  was imposed by the need of fitting the theoretical values of $f$
to the empirical ones. It is also worth mentioning that with the NLTE model atmospheres we got much better agreement.

It should also be highlighted that the opacity modifications can have serious consequences on the stellar structure.
In this paper, we discussed an enhancement of the efficiency of convection in the Z-bump if the opacity is increased.
In turn, more efficient convection affects the mode instability and the flux eigenfunction.

In addition, we found that in some models with the much increased opacity around the Z-bump an additional potential barrier can be formed.
Consequently, the frequencies of the two consecutive radial modes can come close to each other and undergo the avoided-crossing
phenomenon considered up to now only for nonradial modes in main sequence models.

Naturally a number of questions arise: To what extent are these opacity modifications realistic or, more precisely, are they in
the right direction? To what extent is the solution general? Can we apply it to other hybrid B-type pulsators?
A somewhat perplexing result is the need to lower the opacity at $\log T=5.06$ whereas the new local maximum of $\kappa(T)$
has been identified in Kurucz models at this depth. This bump helped to solve the excitation problem with the low frequency modes
in $\delta$ Scuti star models which have much lower masses. Can we have such differences in the opacity profile between stars?
Or maybe our result is only a consequence of the adopted parametrisation to satisfy the data?
The large factor of increase that was imposed on the opacity at $\log T = 5.47$ is also difficult to justify
since our investigations have been unable to identify any missing transitions that would support such an enhancement.
We will be able to try to answer all of these questions if more pulsating stars are analysed in the same way.
We plan such complex seismic studies for the three other main sequence B-type pulsator which show oscillation spectra
of a dual character: $\gamma$ Pegasi, 12 Lacertae, $\alpha$ Lupi.
A need for such significant modifications of stellar opacity can result either from non-homogenous distribution of chemical elements
or from the present-day methods of the opacity computations.
Our main goal was to present a new approach to the analysis of the hybrid pulsators,
which, besides complex seismic modelling, involves the fitting of the mean opacity profile.
Such analysis can lead to important constraints on opacities under the conditions of stellar interiors.

\section*{Acknowledgments}
This work was financially supported by the Polish NCN grants 2015/17/B/ST9/02082.
PW's work was supported by the European Community's Seventh Framework Program (FP7/2007-2013) under grant agreement no. 269194.
Calculations have been partly carried out using resources provided by Wroclaw Centre for
Networking and Supercomputing (http://www.wcss.pl), grant No. 265.
The Los Alamos National Laboratory is operated by Los Alamos National Security, LLC,
for the NNSA of the US DOE under contract number DE-AC5206NA25396.

\label{lastpage}

\end{document}